\documentclass[%
 reprint,
superscriptaddress,
 amsmath,amssymb,
 aps,
]{revtex4-1}

\usepackage{graphicx}
\usepackage{caption}
\usepackage{subcaption}
\usepackage{dcolumn}
\usepackage{bm}
\usepackage{hyperref}
\usepackage{natbib}

\DeclareMathOperator{\Tr}{Tr}

\begin{document}

\title{Tracking Particles with Large Displacements using Energy Minimization}


\author{Rostislav Boltyanskiy}
\affiliation{
Department of Physics, Yale University, New Haven, Connecticut 06520, USA}%

\author{Jason W. Merrill}%
\affiliation{
Desmos, Inc.,  San Francisco, California 94103, USA}%

\author{Eric R. Dufresne}
\affiliation{Department of Materials, ETH Z\"{u}rich, 8092 Z\"{u}rich, CH. }%

\date{\today}

\begin{abstract}
We describe a method to track particles undergoing large displacements. 
Starting with a list of particle positions sampled at different time points, we assign particle identities by minimizing the sum across all particles of the trace of the square of the strain tensor. 
This method of tracking corresponds to minimizing the stored energy in an elastic solid or the dissipated energy in a viscous fluid. 
Our energy-minimizing approach extends the advantages of particle tracking to situations where particle imaging velocimetry and digital imaging correlation are typically required.
This approach is much more reliable than the standard squared-displacement minimizing approach for spatially-correlated displacements that are larger than the typical interparticle spacing.
Thus, it is suitable for particles embedded in a material undergoing large deformations. 
On the other hand, squared-displacement minimization is more effective for particles undergoing uncorrelated random motion.
In the Supplement, we include a flexible MATLAB particle tracker that implements either approach.
This implementation returns an estimation of the strain tensor for each particle, in addition to its identification. 

\end{abstract}

\pacs{Valid PACS appear here}
\maketitle


\section*{Introduction}

In a wide range of pure and applied sciences, the motion of objects needs to be tracked over time.   
The central difficulty is that while an object's trajectory is continuous through space and time, its position can only be sampled at discrete moments.  
When the objects of interest are far away from each other or otherwise distinguishable, tracking is simple.
However, when many identical objects (which we refer to as \emph{particles}) are near each other, tracking can be difficult.

Particle tracking is employed across physics, biology, and many other fields.  
In soft matter physics, it has shed light on mechanical properties like rheology \cite{squires2010}, and revealed microscopic processes underlying phase transitions \cite{peng2011}, among many other applications.
In biology, particle tracking  is used across many length scales from the movement of single molecules, to the transport of organelles within cells, to movement of whole organisms such as flies, birds, fish, and humans \cite{ardekani2013,straw2011,katz2011,rodriguez2011}.

In the absence of Brownian motion, particles embedded in a material reveal its deformation field.
In fluid mechanics, particle tracking reveals flow fields in the Lagrangian frame of reference (\cite{xu2006}). 
Traction force microscopy (TFM) quantifies the forces applied to a solid surface by tracking and analyzing motion of particles embedded within the solid.
TFM was originally developed to quantify the forces exerted by adherent cells (\cite{munevar2001}), but has recently been applied to a wide range of problems in biology and physics  \cite{style2014}.

The standard method of particle tracking assigns identities by minimizing the sum of squared displacements across time points \cite{crocker1996}.  
While this method is rigorously correct for objects undergoing Brownian motion, it works very well across a wide range of applications, provided that the particles move a distance that is small compared to the typical inter-particle separation.
The improvement of this basic tracking approach has been an active area of research in recent years, driven by applications in the biomedical community.  
For a useful comparison of these particle tracking approaches, see \cite{Chenouard2014}.
When the displacements are large, it is helpful to employ a tracking algorithm that exploits knowledge of the system's kinematics.
For example, particle tracking in dense turbulent fluid flows has been greatly improved by employing a ``predictive tracker," which exploits the inertial character of high Reynolds number fluid flow \cite{Ouellette2006}.  

Particle Imaging Velocimetry (PIV) and Digital Imaging Correlation (DIC) are widely used to assess the flow of fluids and deformation of solids \cite{adrian2010,pan2009,Bar-Kochba2015}.
These methods assign velocities or displacements  to locations in a material by cross-correlating patches of an image across time points.
This is a robust and successful approach, even for systems with large displacements.
However, it is inappropriate in cases where individual particle identities need to be followed over time or where the displacements need to be known at the resolution of individual particles.

Here, we extend particle tracking to the large-displacement regime where PIV and DIC are usually employed.
Our algorithm is optimized for tracking the motion of particles undergoing large spatially-correlated displacements, typical for tracers embedded in a deformed elastic solid or flowing viscous fluid. 
We start by posing the tracking problem generally, and review the maximum likelihood method employed to track Brownian particles.
Then, we describe our energy minimization approach and directly compare the results of the two methods for simulated and real data.

\section*{Optimal Assignment}
\vspace{2mm}

Suppose a set of particles, $\{p^{(k)}(t_{i})\}$, where $k$ is the particle index, are found at locations $\{\vec{x}^{(k)}(t_{i})\}$ at time $t_{i}$ and a set of particles $\{p^{(k)}(t_{f})\}$ are found at $\{\vec{x}^{(k)}(t_{f})\}$ at time $t_{f}$.
We wish to connect particle positions across time to make trajectories.
The particle identities are represented by a list, $T$, where $T_{l}=m$ if particle $p^{(l)}(t_{i})$ at time $t_{i}$ becomes particle $p^{(m)}(t_{f})$ at time $t_{f}$.  

\begin{figure}
\centerline{\includegraphics[width=.45\textwidth]{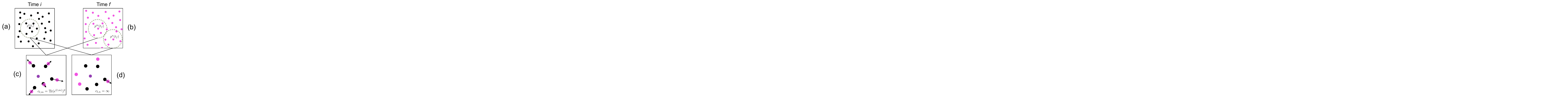}}
\caption{Schematic of strain-based particle tracking. (a)  Particles at time $t_{i}$. Dotted circle represents a region around the particle of interest, $p^{(l)}(t_{i})$, with radius $r_{max}$ within which the particle neighbors are considered.  (b) Particles at time $t_{f}$ with candidate particles circled as in (a).  (c)  Particle  $p^{(l)}(t_{i})$ and its neighbors overlapped with $p^{(m)}(t_{f})$ and its neighbors.  Arrows show tracked displacements between neighbors of the candidate particles.  (d) Particle $p^{(l)}(t_{i})$ and its neighbors overlapped with $p^{(n)}(t_{f})$ and its neighbors.  Arrows show tracked displacements between neighbors of the candidate particles.}
\end{figure}

We find an optimal assignment of identities by minimizing a cost function.
We associate a cost, $c_{l, m}$, to each possible particle pairing.  
The total cost, $C$, is the sum of the costs of individual pairings across all of the particles, $C=\sum_{l}  c_{l, T(l)}$.  
We assign identities that minimize the cost using the Hungarian algorithm. \cite{Kuhn2005, Buehren2004}.  

Since assigning particles is a combinatoric process, it rapidly becomes computationally overwhelming as the system size increases. 
To avoid this difficulty, it is essential to limit the number of combinations by ruling out unphysical assignments, as described below.

\section*{Minimization of Squared-Displacements}
The standard particle-tracking algorithm minimizes the sum of the squared displacements across all of the particles \cite{crocker1996}.
This yields the most likely assignment of identities when particles undergo Brownian motion.
In that case, the probability that the $k^{th}$ particle will be displaced by $\Delta \vec{x}^{(k)}$ in a time interval, $\Delta t$, is: 

\begin{equation}
P(\Delta \vec{x}^{(k)}, \Delta t) \sim \exp \left(
-\frac{\left|\Delta \vec{x}^{(k)}\right|^2}{4dD\Delta t} 
\right) \\
\end{equation}

Here, $D$ is the diffusion coefficient and $d$ is the number of spatial dimensions \cite{berg1993}. 
Therefore, the probability of observing a specific set of displacements of $N$ identical particles moving independently is:

\begin{eqnarray}
P &=& \prod\limits_{k=1}^{N} P(\Delta \vec{x}^{(k)}, \Delta t) \nonumber \\
   &\sim& \exp \left( -\sum\limits_{k=1}^{N} \frac{\left|\Delta \vec{x}^{(k)}\right|^2}{4dD\Delta t} \right)
\end{eqnarray}

The most likely assignment of particle identities maximizes the total probability, $P$, which is equivalent to minimizing the total squared-displacement  $\sum\limits_{k=1}^{N} \left|\Delta \vec{x}^{(k)}\right|^2$.   
Furthermore, we can assign costs for potential pairings as $c_{l,m} = |\vec{x}^{(l)}(t_{i}) - \vec{x}^{(m)}(t_{f})|^2$, and minimize the cost function $C$, as defined in the previous section.  

To accelerate the data analysis, one needs to rule out unphysical assignments. 
Typically, one specifies a maximum displacement a particle could have between time points, and assigns an infinite cost to all of the elements of $c$ that correspond to displacements that would exceed this maximum.
These elements are ignored in the combinatorial optimization.
  
In practice, minimizing the squared-distance works very well whenever particle displacements are small compared to the distance of a particle to its nearest neighbors.


\section*{Minimization of Energy}
\vspace{2mm}

In the case of large displacements, square-distance minimizing trackers may not work effectively.  
In many cases, however, these displacements may have strong spatial correlations.
For example, particles embedded in a rigid material undergoing translation and rotation are fixed relative to their neighbors. 
Similarly, displacements of particles embedded in a liquid or solid undergoing large deformations are typically strongly correlated to their neighbors.
In both these cases, we do not wish to penalize displacements in the laboratory frame, but displacements relative to neighboring particles.

In continuum mechanics, the variation of displacements over space is characterized by the strain.
In the case of a linear, isotropic, elastic solid of Young's modulus, $E$,  the stored energy is
\begin{equation}
U_{el} = \frac{E}{2}\int_{vol}  \Tr \epsilon^2 dV
\end{equation}
where $\epsilon$ and $\vec{u}$ are the strain and displacement fields, respectively.
In component form, these two are related as $\epsilon_{ij} = \frac{1}{2} (\partial_{i} u_{j} + \partial_{j} u_{i})$ \cite{landau1986}.
Analogously, for a fluid of dynamic viscosity $\mu$ sheared with a strain rate $\dot \epsilon$, the rate of energy dissipation is \cite{kundu2008}:
\begin{equation}
\dot U_{fl} = 2 \mu \int_{vol}  \Tr \dot \epsilon^2 dV
\end{equation}
If one is considering only two time points, then experimental estimates of $\epsilon$ and $\dot \epsilon$ only differ by a pre-factor.

Motivated by these two physical examples, we propose to assign particle identities by minimizing the stored/dissipated energy.
The basic hypothesis is that tracking errors tend to exaggerate the strain/strain-rate, increasing the apparent stored/dissipated energy. 
Therefore, we implement a cost function $c_{l,m} = \Tr (\epsilon^{(l,m)})^2$.  
The main challenge of the proposed algorithm is to identify the strain associated with a given particle pairing.  
Calculating the strain requires positions of particles and their nearest neighbors at two times.

The nearest neighbors of particle $p^{(k)}$, can be found in 2-D using the Delaunay triangulation.
To ensure that the strain is uniform over the region where it is calculated, we only consider neighbors within a distance $r_{max}$ of $p^{(k)}$.
In Fig. 1(a,b), neighbors included in the strain calculation are within the dotted circles.

We use the method of Falk and Langer \cite{Falk1998} to calculate the strain associated with candidate particle pairs and their neighbors.  
The strain about a particle at position $\vec{x}$ from the position of neighbors $\vec{x}^{(n)}$ can be estimated as follows:

\begin{eqnarray}
X_{ij} &=& \sum_{n} (x_{i}^{(n)}(t_{f}) - x_{i}(t_{f})) \cdot (x_{j}^{(n)}(t_{i}) - x_{j}(t_{i})), \\
Y_{ij} &=& \sum_{n} (x_{i}^{(n)}(t_{i}) - x_{i}(t_{i})) \cdot (x_{j}^{(n)}(t_{i}) - x_{j}(t_{i})), \\
\Lambda_{ij} &=& X_{ik}Y_{jk}^{-1} ,\\
\label{eq:strain}
\epsilon_{ij} &=& \frac{1}{2} (\Lambda_{ij} + \Lambda_{ji}) - \delta_{ij}.
\end{eqnarray}
Note that in order to calculate strain accurately, at least $d$ neighbors must be included for particle tracking in $d$-dimensions.

To properly calculate strain around a particle, one needs to accurately track its neighbors across time. 
Consider a particle at time $t_{i}$, $p^{(l)}(t_{i})$ (in the center of the dotted circle in Fig. 1a) and a candidate corresponding particle at time $t_{f}$, $p^{(m)}(t_{f})$ (in the center a dotted circle in Fig. 1b).  
To connect the neighbors of $p^{(l)}(t_{i})$ with those of $p^{(m)}(t_{f})$, we translate the candidate particles and their neighbors to the same location, as shown in Figure 1(c,d). 
We then track the neighbors of the candidate pair by minimizing the square of the residual displacements. 
To accelerate the calculation, we rule out relative displacements that would exceed an expected maximum value of the strain.
If the candidate pair produces at least $d$ tracked neighbors, we calculate  the strain for the candidate pairing according to Eq. (\ref{eq:strain}) and assign the pairing a cost $c_{l,m} = \Tr (\epsilon^{(l,m)})^2$.
Otherwise, the pairing is ruled out by setting the cost to infinity.
Once all possible pairings have been assigned costs, the Hungarian algorithm is used to minimize the total cost across all of the particles.  

\section*{Implementation of tracking algorithms}
We implement the squared-distance-minimizing ``diffusion tracker" and energy-minimizing ``strain tracker" algorithms in MATLAB.
In the supplement, we include the main particle-tracking function, \emph{Tracker.m}.
We also include a script, \emph{Example.m}, that allows the user to explore the examples described below. 
Details regarding use of the code are found in the comments as well as the text of \emph{ReadMe.doc}.
A convenient by-product of our tracker is that it returns the symmetrized strain matrix for each tracked particle.

\section*{Comparison of tracking strategies with simulated data}

\begin{figure*}
\centerline{\includegraphics[width=1\textwidth]{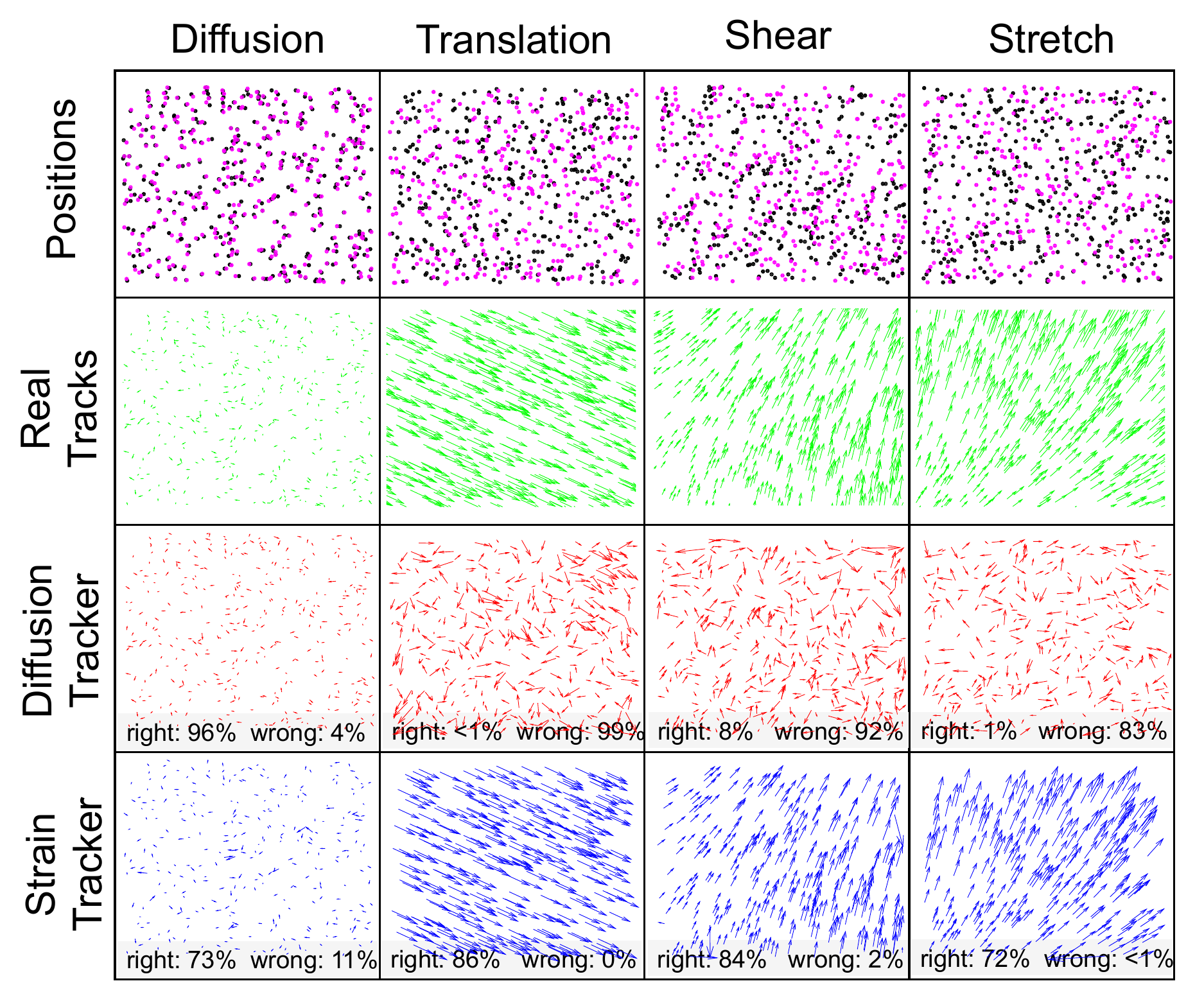}}
\caption{Comparison of the diffusion and strain trackers for  simulated data of particles moving by diffusion (first column), translation (second column), shear (third column), and stretch (fourth column).  The particle positions at the first time point (black dots) and second time point (magenta dots) are shown in the first (top) row. The correct displacments (green arrows) are shown in the second row.  The displacments calculated from identities returned by the diffusion and strain trackers are shown in the third and fourth rows, respectively.   In all cases, arrows are drawn to scale. In the third and fourth rows, the numbers in gray boxes correspond to percentages of particle trajectories calculated correctly and incorrectly by each tracker. The percentages do not add to 100\% because the trackers were not able to find partners for some particles.}
\end{figure*}

We compare the performance of the diffusion and strain trackers with simulated data.  
As shown in Figure 2,  we consider four types of particle motion: diffusion, translation, shear, and stretch.
For each example, the first row of Figure 2 shows the positions of particles at the first (black) and second (magenta) time points.
The second row shows the displacements for correctly tracked particles in green.
The third and fourth rows show the displacements determined by the diffusion tracker (red) and the strain tracker (blue).

\begin{figure*}
\centerline{\includegraphics[width=1\textwidth]{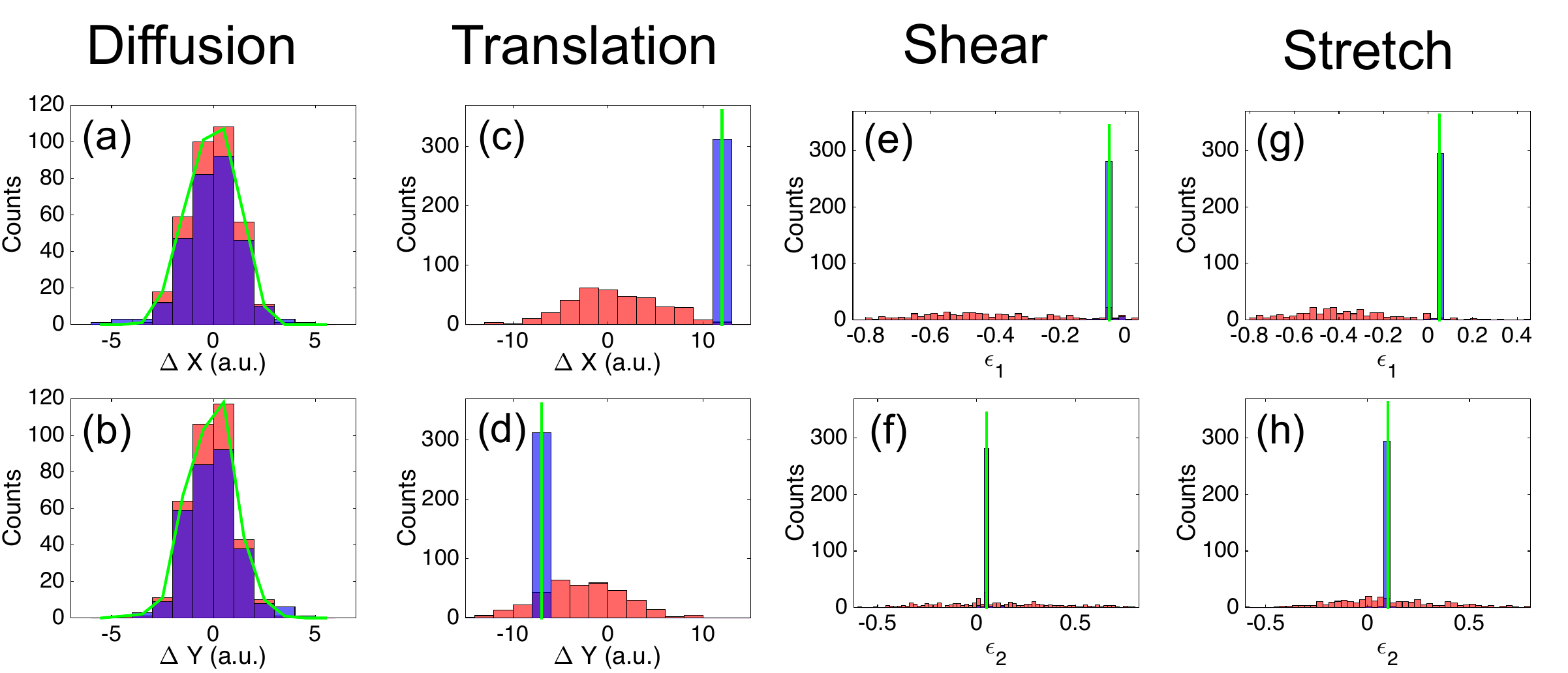}}
\caption{Histograms of individual particle displacements (a-d) and strain eigenvalues (e-h) from analysis of simulated data in Figure 2. (a) and (b) are histograms of horizontal and vertical displacements, respectively, from simulated diffusion data. (c) and (d) are histograms of horizontal and vertical displacements, respectively, from simulated translation data. (e) and (f) are histograms of strain eignevalues from a simulated shear. (g) and (h) are histograms of strain eigenvalues from a simulated stretch. In all panels, results from the diffusion tracker are in red, those of the strain tracker are in blue, and the overlap is in purple. In green is an outline of the histogram of correct displacements in (a),(b) and a vertical line corresponding to the correct values of displacements and strains in (c)-(h).}
\end{figure*}

Not surprisingly, the diffusion tracker out-performs the strain tracker for the case of diffusion. 
For the example shown in Figure 2, the diffusion tracker returns results for all of the particles.  
About 96\% of these tracks are correct.
On the other hand, the strain tracker provides identifications for only 84\% of the particles, and about 15\% of these are incorrect.
These tracking errors impact quantitative measures of the particle motion. 
Even though the distributions of the particle displacements for the two trackers look similar,  Figure 3, tracking errors  tend to introduce counts in the tails of the displacement distribution that have a significant impact on the mean-squared particle displacement (MSD).
In this example, the MSD calculated from the diffusion tracker is within 2\% of the correct value.
However, the value calculated using the strain tracker is 30\% greater than the correct value.

On the other hand, for the cases of large displacements due to translation, shear, and stretch, the strain tracker is a much more reliable choice.
For the examples in Figure 2, the diffusion tracker returns tracks for greater than 84\% of the particles, but less than 8\% of them are correct.
On the other hand, the strain tracker identifies greater than 73\%  of the particles and less than 3\% of them are incorrect.
Furthermore, for the case of the strain tracker, dropped particles and errors tend to be localized to the boundaries of the field of view, which are easily discarded when greater accuracy is required.

In these cases, the strain tracker accurately quanitfies particle displacements and strains.
The expected displacements and strains for the cases of translation, shear, and stretch are shown as vertical green lines in Figure 3.
The red histograms display the values for each particle returned by the diffusion tracker and the blue histograms report those from the strain tracker.
The strain tracker consistently returns the correct values, while the incorrect tracks from the diffusion tracker return broadly distributed incorrect values.

\section*{Comparison of tracking strategies with experimental data}

In this section, we compare the diffusion and strain tracker with some experimental data. 

\begin{figure}
\centerline{\includegraphics[width=.5\textwidth]{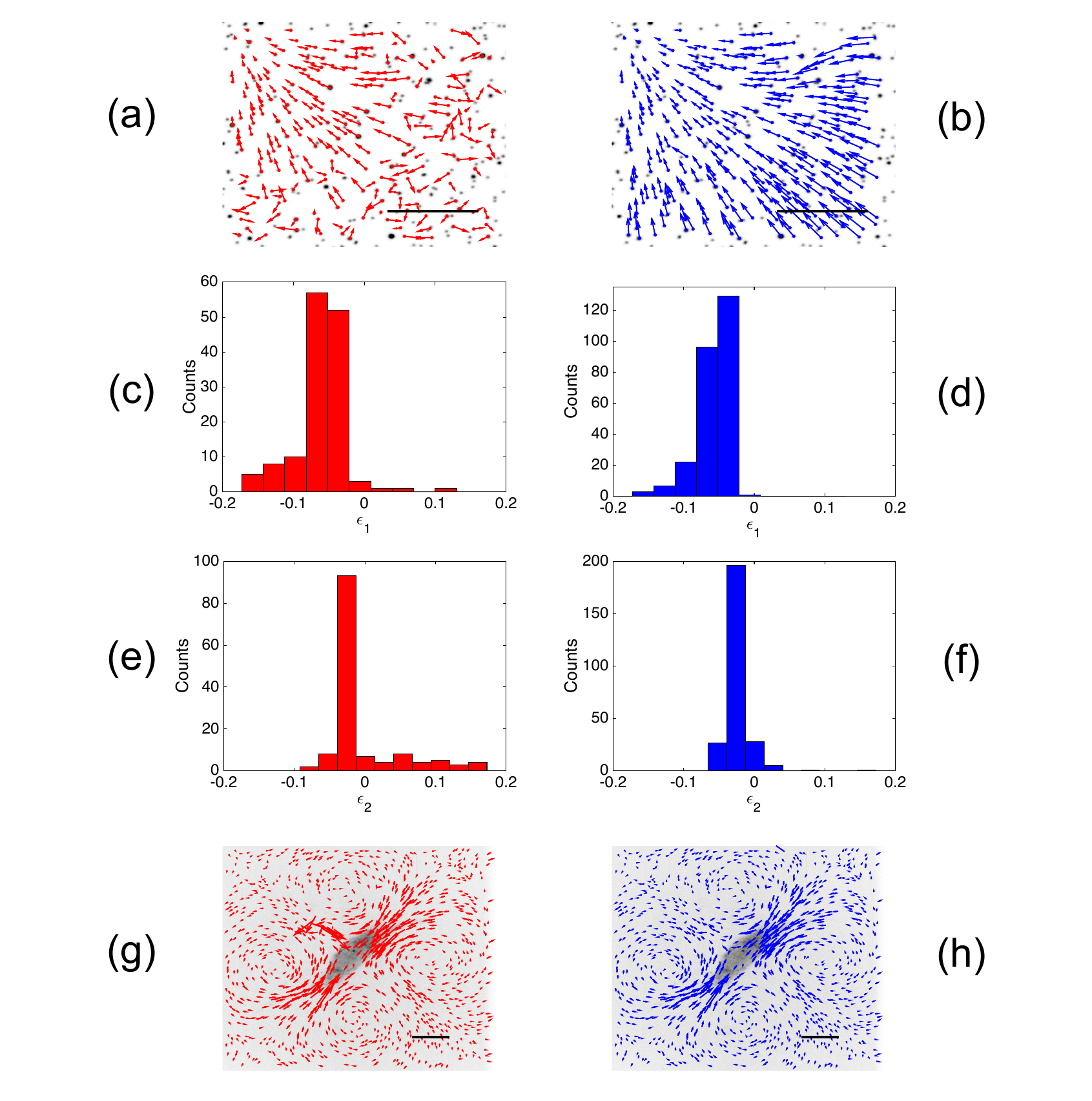}}
\caption{Particle tracking examples with experimental data.\\ (a)-(b) Trajectories of particles embedded in a silicone gel undergoing uniform compression, calculated with the diffusion tracker (a) and the strain tracker (b). Arrows are scaled by a factor of 2. (c)-(f) Histograms of strain eigenvalues as calculated with the diffusion tracker (c), (e) and the strain tracker (d),(f). (g)-(h) A contractile fibroblast cell adherent on a silicone gel.  Arrows overlaid on top represent particle displacements from cell traction forces as calculated with the diffusion tracker (g), and the strain tracker (h).  Arrows are scaled by a factor of 5. 
Scale bars are $20\mu m$}
\end{figure}

First, we assess both trackers' ability to identify particles during a relatively large and homogeneous strain.
Here, we deform a silicone gel with embedded fluorescent tracers using a macroscopic deformation. 
We image the tracers over a small field of view, where we expect a reasonably uniform compression.
We manually applied a uniform affine transformation to the particle locations to estimate the strain, resulting in eigenvalues of -0.06 and -0.04.
Then,we calculated particle displacements with the strain and diffusion trackers, as shown in Figure 4(a,b).
While both trackers return comparable results where the displacements are small (upper-left corner of the field of view), they disagree where the displacements are large.
In this region, the diffusion tracker returns an uncorrelated displacement field, while the strain tracker returns the expected compression.
Our strain tracker returns a strain tensor associated with the displacement of each tracked particle.
Histograms of the strain eigenvalues of each particle are plotted in Figure 4 c-f.
For both trackers, the peaks of the histogram agree well with the expected values from the manual affine transformation.
However, the diffusion tracker reports a much broader distribution, including some unphysical positive values. 

Next, we consider an example from Traction Force Microscopy (TFM) where the strains have a much stronger spatial heterogeneity.
In TFM, forces exerted by small objects, such as cells, are quantified by measuring the deformation of an elastic material they are adhered to. 
To quantify the deformation, fluorescent particles are embedded in the elastic material.
Displacements caused by a fibroblast cell adherent to a silicone gel are presented in Figure 4 g,h.
The cell is fluorescently tagged and is displayed in inverted contrast.
Overlaid on top of the cell image are displacements of the surface underneath the cell (scaled by a factor of 5), determined by the (g) the diffusion tracker and (h) strain tracker.
While displacements measured by the strain tracker and  diffusion tracker mostly agree, the diffusion tracker identifies some unexpected large strains near the center of the cell.

\section*{Conclusion}

We have introduced a particle tracking algorithm based on the minimization of energy.
This approach out-performs the conventional squared-displacement minimizing particle tracker when the displacements are larger than the interparticle spacing.
On the other hand, our
strain tracker may have difficulty when  the strain changes significantly over the typical interparticle spacing.
This can occur in the vicinity of strain singularities, such as cracks, 
and when particles are moving randomly, as in Brownian motion.
Although, the code included in the supplement is designed for two time points and two spatial dimensions, expanding to three dimensions and many time points is a straightforward modification.

\section*{Acknowledgments}
We thank Madhusudhan Venkadesan, Katharine Jensen, and Robert Style for helpful discussions and support from the Army Research Office Multidisciplinary University Research Initiative (ARO MURI) W911NF-14-1-0403.

%

\end{document}